\documentclass{appolb}
\usepackage{epsfig}
\usepackage{amssymb}
\usepackage{amsmath}
\usepackage[cp1250]{inputenc}

\begin{document}
\title{Exotic Smoothness and Astrophysics}
\author{Jan S\l adkowski
\address{Institute of Physics, University of Silesia\\ ul. Uniwersytecka 4, Pl 40007 Katowice, Poland}
}
\maketitle
\begin{abstract}
The problem of possible astrophysical consequences
of the existence of exotic differential structures on manifolds is discussed. It is argued that corrections to the curvature of the form of a source like terms should be expected in the Einstein equations if they are written in the "wrong" differential structure.
Examples of topologically trivial spaces on which
exotic differential structures act as a source of gravitational force even in
the absence of matter are given. Propagation of light in the presence of such phenomena is also discussed. A brief review of exotic smoothness is added for completeness.
\end{abstract}
\def\R{\mathbb{R}}
\section{Introduction}
In 1854 B. Riemann, the father of contemporary differential geometry,  suggested that
the geometry of space
may be more than just a mathematical tool defining a stage for
physical phenomena, and may in fact have profound physical meaning
in its own right \cite{[1}. But is it reasonable to contemplate {\it to what extent the choice of mathematical model for spacetime has important
physical significance}? With the advent of general relativity
physicists began to think of the spacetime as a differential
manifold. Since then various assumptions about
the spacetime topology and geometry have been put forward. But why should the choice of differential structure of the spacetime manifold matter?  Most of topological spaces used
for modelling spacetime have natural differential structures and the question of their non-uniqueness seemed to be extravagant.
Therefore,
the counterintuitive discovery of exotic four dimensional Euclidean
spaces following
from the work  of Freedman \cite{[3} and Donaldson \cite{[4} raised various
discussions about the possible physical consequences of this discovery \cite{[6}-\cite{[24}. It has been shown that exotic
(nonunique) smooth structures are
especially abundant in dimension four \cite{[18}  and there is at least a two parameter family of exotic ${\R}^{4}$'s. Such manifolds play important r\^ ole in
theoretical physics and astrophysics and it became necessary to investigate
the physical meaning of exotic smoothness.
Unfortunately, this is not an easy task: we only know
few complicated coordinate descriptions \cite{[19} and most
mathematicians believe
that in most cases there there might not  be any finite atlas on an exotic ${\R}^{4}$ and other
exotic four-manifolds (three coordinate patches description seems to be the best achievement). Since their discovery, exotic ${\R}^{4}$ have revealed themselves in various physical contexts. For example, some non-perturbative limits of a QCD-like YM theory can be reached when
the theory is formulated on 4-manifolds which are locally exotic ${\R}^{4}$; 
exotic ${\R}^{4}$'s in a region of the spacetime can act as the sources of the
magnetic field in this spacetime and imply electric charge quantization;  calculations on exotic ${\R}^{4}$'s can be formulated  in covariant
cohomologies (sheaves and gerbes on goupoids) or in 2-dimensional
quantum CFT and in that way imply quantum gravity  corrections if compared
to the calculations done in the standard ${\R}^{4}$. Such phenomena are
counterintuitive but I am  aware of no physical principle that would require
rejection of such spacetimes or solutions.
\section{Astrophysical consequences of the existence of exotic smoothness}
\subsection{Nonequivalent differential structures}
It is possible that two manifolds $M_1$ and $M_2$ are homeomorphic but not diffeomorphic, that is they are identical as topological spaces ($M$) but no bijection (1-1 mapping) between them is differentiable\footnote{To be precise, these functions must not be differentiable at least one point.}. In that case, we say that the underlying topological space $M$ has nonequivalent differential structures often referred to as exotic differential structures. Therefore, exotic ${\R}^{4}$'s
are defined as four-manifolds that are homeomorphic to the fourdimensional
Euclidean space ${\R}^{4}$ but not diffeomorphic to it.  The existence of nonequivalent
differential structures does not change the definition
of the derivative. The essential difference is that the  sets (actually algebras) of real
differentiable functions are different on non-diffeomorphic manifolds. In
the case of exotic ${\R}^{4}$'s this means that there are
continuous functions  ${\R}^{4} \mapsto {\R}$ that are smooth
on one exotic ${\R}^{4}$ and only continuous on another and vice
versa \cite{[20}.
\subsection{General relativity on exotic ${\R}^{4}$'s with few symmetries}
 Suppose we are given an exotic ${\R}_{\theta}^{4}$
with few symmetries\footnote{ We say that a smooth
manifold has few symmetries provided that for every choice of differentiable
metric tensor, the isometry group is finite.}. We can try to solve the
Einstein equations on this ${\R}_{\theta}^{4}$. Suppose we have found such a
solution. Whatever the
boundary conditions be we would face one of the two following
situations \cite{[24}.
\begin{description}
  \item[\checkmark]\ \ The isometry group G of the solution  acts properly\footnote{We say that
$G$  acts properly on $X$ if and only if for all compact subsets $Y
\subset X$, the set $\{ g\in G:gY\cap Y\not= \emptyset \}$ is also
compact.} on
${\R}_{\theta}^{4}$. Then  G is finite. There is no
nontrivial Killing vector field and no solution to Einstein equations can be stationary. The gravity is quite "complicated" and even empty spaces do evolve.
\item[\checkmark]\ \ The isometry group G of the solution  acts nonproperly on
${\R}_{\theta}^{4}$. Then G is locally isomorphic to SO(n,1) or SO(n,2 ).
But the nonproper action of G on ${\R}_{\theta}^{4}$ means that there are
points infinitely close together in ${\R}_{\theta}^{4}$ ($x_{n} \rightarrow
x$) such that arbitrary large different isometries ($g_{n} \rightarrow
\infty $) in G maps them into infinitely close points in ${\R}_{\theta}^{4}$
($g_{n}x_{n} \rightarrow y \in {\R}_{\theta}^{4}$). There must exists quite
strong gravity centers to force such convergence (even in empty spacetimes).
\end{description}
 We see that in both cases Einstein gravity is quite
nontrivial even in the absence
of matter. Recall that if a spacetime has a Killing vector field
$\zeta ^{a}$, then every covering manifold admits appropriate Killing vector
field $\zeta ^{'a}$ such that it is projected onto $\zeta ^{a}$ by the
differential of the covering map. This means that discussed above
properties are
inherited by  any space that has exotic ${\R}^{4}$ with few symmetries as
a covering manifold e.g. quotient manifolds obtained by a smooth action
of some finite group.
\subsection{String-like gravitational sources}
  Asselmeyer considered a topological manifold $M$ that can be given two inequivalent differential structures  $M'$ and
$M''$ and found the change in covariant
derivative induced by exoticness \cite{[21}. Consider a 1-1 map ${\alpha}:\  M' \rightarrow M''$ that is not
a diffeomorphism at some point $p_{0} \in M'$.
The splitting of the map $d\alpha : \ TM' \rightarrow TM''$
in some neighborhood $U(p_{0})$ of the point $p_{0}$:
$$ d\alpha \mid _{U(p_{0})} = \left( b_{1},b_{2} \right) $$
allows us to express  the change in the covariant derivative in the following form:
$$ \nabla ''=\nabla '+  \left( b_{1}^{-1}db_{1} \right) \oplus
\left( b_{2}^{-1}db_{2} \right). $$
This made it possible to calculate the
corresponding change in
the curvature tensor and Einstein equations. Recall that the curvature tensor is given by $$R\left( X,Y\right)Z = \nabla _{X}\nabla _{Y} Z-\nabla _{Y}
\nabla _{X} Z + \nabla _{\left[ X,Y\right] } Z, $$
where $X,\ Y,\ Z$ are vector fields.  This
means that we should expect that
$$ Ric \left( X, Y\right) -\frac{1}{2} g\left( X, Y\right) R\not= 0
 $$
in $M''$ even if the right hand side vanishes on $M'$! Asselmeyer argues for a string-like interpretation of this source term.
  Suppose we have discovered some strange
 astrophysical source of gravitation that
do not fit to any acceptable
solution of the Einstein equations\footnote{For example, a cosmic string-like lensing is possible. A priori, it should be possible to distinguish between standard cosmic string and Asselmeyer's strings  because cosmic strings forms conic spacetimes with rather trivial gravity.}. This may mean that we are using
wrong differential structure on the spacetime manifold and this strange
source is sort of an artefact of this mistake. If we change the
differential structure  then everything would be OK.
\subsection{Why should exotic structures  matter?}
A manifold $M$ can be equivalently described by its algebra of real differentiable functions $C(M,\R)$. A. Connes managed to generalize
this result for a much larger class of algebras, not
necessarily commutative \cite{con}. The idea behind this is that one have to define the appropriate Dirac operator and the differential calculus is recovered by commutators of functions with the Dirac operator. This mean that the spacetime structure is  actually given by those properties of matter fields that are governed by Dirac equations. It is possible that there are some subtleties in fundamental interaction that reveal themselves only on astrophysical scales and exotic differential structures might be necessary to take them into account.
\section{Astrophysical observations of exotic smoothness}
\subsection{Maxwell equations in gravitational background}
{ Maxwell's equations in the background metric
in empty space can be written as \cite{[25}\footnote{There still are controversies concerning definition of electromagnetic field in gravitational background but such details are unimportant here.}:}
\begin{alignat*}{4}
\nabla\cdot \textbf{B} &= 0, \qquad &\nabla\times \textbf{H} - \frac{1}{c\sqrt{\gamma}}\frac{\partial}{\partial t}\sqrt{\gamma}\textbf{D}-\frac{4\pi}{c} \textbf{s}&= 0\\
 \nabla \cdot\textbf{D} &= 0,  &\nabla\times \textbf{E} +\frac{1}{c\sqrt{\gamma}}\frac{\partial}{\partial t}\sqrt{\gamma}\textbf{B}&= 0
 \notag ,
\end{alignat*}
 where $\gamma =det\ g_{ik}$ \hspace{.2cm} $\textbf{s}^{i}=\rho \frac{dx^{i}}{dt}$. The fields \textbf{B} and \textbf{D}  should be modified as:
\begin{alignat*}{4}
 \textbf{D} &=\frac{ \textbf{E}}{\sqrt{h}}+\textbf{H}\times \textbf{G}, \qquad & \textbf{B} &= \frac{\textbf{H}}{\sqrt{h}}+ \textbf{G}\times \textbf{E}
 \notag
\end{alignat*}
with $h=g_{00}$  and $\textbf{G}_{i}=-\frac{g_{0i}}{g_{00}}.$
Maxwell's equations in the background metric
in empty space can be written as:
\begin{alignat*}{4}
\nabla\cdot \textbf{B} &= 0, \qquad &\nabla\times \textbf{H} - \frac{1}{c}\frac{\partial}{\partial t}\textbf{D}&= 0\\
 \nabla \cdot\textbf{D} &= 0,  &\nabla\times \textbf{E} +\frac{1}{c}\frac{\partial}{\partial t}\textbf{B}&= 0
 \notag .
\end{alignat*}
Simple calculations \cite{[26} shows that one should expect modification of the dispersion relation of the form:
$$ \textbf{k}^2 -\omega ^2 -2\textbf{G}\cdot \textbf{k}\omega =0.$$
This corresponds to a subluminal propagation of electromagnetic radiation. Such effects are observable but without explicit solution for the metric tensor it would be difficult to ascribe them to the exoticness.
\subsection{Non-gravitational effects and observations?}
Every measurement performed by humans involves gauge interactions (\eg electromagnetic interaction). By using the heat kernel method
\cite{[33} we can express the Yang-Mills action in the form \cite{[34}:
$${\cal L}_{YM}\left( F\right) \sim \lim _{t\to 0}\frac{tr\left( F^{2}
exp\left( -t{\cal D}^{2}\right) \right)}{tr\left( exp \left(-t{\cal D}^{2}
\right) \right) } \ ,$$ where ${\cal D}$ is the appropriate Dirac operator.
Suppose that we have a one parameter ($z$) family of differential structures
and the corresponding family of Dirac operators ${\cal D}(z)$. The Duhamels's
formula
$$\partial _{z} \left( e^{-t{\cal D}^{2} \left( z\right) }\right) =
\int _{0}^{t} e^{-\left( t-s\right) \triangle \left( z\right) } \partial _{z}
\left( {\cal D}^{2} \left( z\right) \right)
e^{-s \triangle \left( z\right) }ds \ ,$$
where $\triangle $ is the scalar Laplacian, can be used to calculate the
possible variation of ${\cal L}_{YM} (F)$ with respect to $z$.
Unfortunately our present knowledge of exoticness is to poor for
performing such calculations. Nevertheless, we can try to estimate the possible effects in the following way.
For an operator $K$ with a smooth kernel we have the following
asymptotic formula \cite{[35}:
$$ tr \left( Ke^{-t{\cal D}^{2}} \right)  \sim tr \left( K\right) + \sum
^{\infty} _{i=1} t^{i}a_{i}\ ,$$ where the spectral coefficients $a_i$  describe some important details of geometry of the underlying maifold \cite{[js1,[js2}. So if $F^{2}$ is smooth with respect to all differential structures (e.g. has
compact support \cite{[20}) then the possible effects of exoticness might "decouple" (are
negligible). This means that we are unlikely to discover exoticness by
performing "local" experiments involving gauge interactions. If we consider only matter (fermions) coupled to gravity then the action can also 
be expressed in terms of the coefficients of the heat kernel expansion of the
Dirac Laplacian, ${\cal D}^{2}$. In this case we may be able
to detect  the exoticness of differential structures only if the Dirac operator specifies 
it almost uniquely \cite{[9, [35}.
\section{Conclusions}
If exotic smoothness has anything to do
with the physical world it may be a source/ explanation of various
astrophysical and cosmological phenomena. Dark matter, vacuum energy
substitutes and strange attracting/scattering centers are the most obvious among them.  Exoticness of the spacetime might be responsible for the anomalies in  supernovae properties or other high density objects \cite{[27}-\cite{[35}. Further research to distinguish exotic smoothness from other causes is necessary but it is a very nontrivial  task.

\end{document}